\documentclass{optica-article}

\journal{opticajournal} 

\articletype{Research Article}

\begin{document}

\title{Integrated waveguide-based acousto-optic modulation with near-unity conversion efficiency}

\author{Liang Zhang,\authormark{1} Chaohan Cui,\authormark{1} Pao-Kang Chen,\authormark{1} and Linran Fan\authormark{1,*}}

\address{\authormark{1}James C. Wyant College of Optical Sciences, The University of Arizona, Tucson, Arizona 85721, USA}

\email{\authormark{*}lfan@optics.arizona.edu} 


\begin{abstract*} 
Acousto-optic modulation in piezoelectric materials offers the efficient method to bridge electrical and optical signals. It is widely used to control optical frequencies and intensities in modern optical systems including \textit{Q}-switch lasers, ion traps, and optical tweezers. It is also critical for emerging applications such as quantum photonics and non-reciprocal optics. 
Acousto-optic devices have recently been demonstrated with promising performance on integrated platforms. However, the conversion efficiency of optical signals remains low in these integrated devices. This is attributed to the significant challenge in realizing large mode overlap, long interaction length, and high power robustness at the same time.
Here, we develop acousto-optic devices with gallium nitride on sapphire substrate. The unique capability to confine both optical and acoustic fields in sub-wavelength scales without suspended structures allows efficient acousto-optic interactions over long distances under high driving power. This leads to the near-unity optical conversion efficiency with integrated acousto-optic modulators. With the unidirectional phase matching, we also demonstrate the non-reciprocal propagation of optical fields with isolation ratio above 10\;dB.
This work provides a robust and efficient acousto-optic platform, opening new opportunities for optical signal processing, quantum transduction, and non-magnetic optical isolation.

\end{abstract*}

\section{Introduction}
Large-scale integration and device minimization are powerful methods to improve system functionality and efficiency. This is witnessed by the recent development of nonlinear optics in integrated photonic circuits~\cite{kippenberg2018dissipative,marpaung2019integrated,bogaerts2020programmable,elshaari2020hybrid}. Compared with optical fields, acoustic fields have much lower propagation speed and stronger coupling with electric fields, thus providing complementary benefits for signal processing~\cite{safavi2019controlling}. Therefore, tailored interactions between optical and acoustic fields in hybrid photonic-phononic circuits attract significant attentions recently with potential applications ranging from quantum transduction~\cite{bochmann2013nanomechanical,fan2016integrated,satzinger2018quantum} and comb generation~\cite{shao2020integrated,fan2019spectrotemporal} to photonic machine learning~\cite{zhao2022enabling}.


Acousto-optic modulation (AOM) plays a critical role in such hybrid circuits for signal conversion among different degrees of freedom. Intensive efforts have been devoted to the development of integrated acousto-optic modulators based on different piezoelectric materials including lithium niobate~\cite{sarabalis2020acousto, shao2020integrated,sarabalis2021acousto}, aluminum nitride~\cite{liu2019electromechanical,kittlaus2021electrically,tadesse2014sub}, gallium arsenide~\cite{de2006compact}, and indium phosphire~\cite{renosi1993resonant}. Efficient AOM requires the simultaneous confinement of optical and acoustic fields in sub-wavelength structures~\cite{sarabalis2021acousto}. While optical confinement can be readily realized using materials with higher refractive index for waveguides, acoustic confinement is challenging as integrated photonic materials typically have acoustic velocities higher than substrates~\cite{fu2019phononic}. The lack of acoustic confinement leads to small coupling strengths between optical and acoustic fields due to the small mode overlapping. While the simultaneous confinement of optical and acoustic fields can be realized in suspended structures, the interaction length and power handling capability are limited due to the mechanical fragility~\cite{sarabalis2021acousto}. The complex fabrication process also causes high acoustic propagation losses, which further deceases the interaction length.
As a result, high optical conversion efficiency is still out of reach for integrated acousto-optic devices.

In this article, we overcome these challenges to realize near-unity conversion efficiencies with integrated acousto-optic modulators. This is achieved by developing the gallium nitride (GaN) on sapphire platform. The refractive index of GaN is significantly larger than sapphire~\cite{zheng2022integrated}. More importantly, velocities of both transverse and longitudinal acoustic waves in GaN are remarkably lower than sapphire~\cite{fu2019phononic}. Therefore, we can realize sub-wavelength confinement of both optical and acoustic fields in GaN waveguides on sapphire substrates without suspended structures. Strong acousto-optic coupling can be realized over long interaction lengths under high driving power with minimal propagation losses, leading to the near-unity optical conversion efficiency.
\begin{figure*}[tbp]
\centering\includegraphics[width=\textwidth]{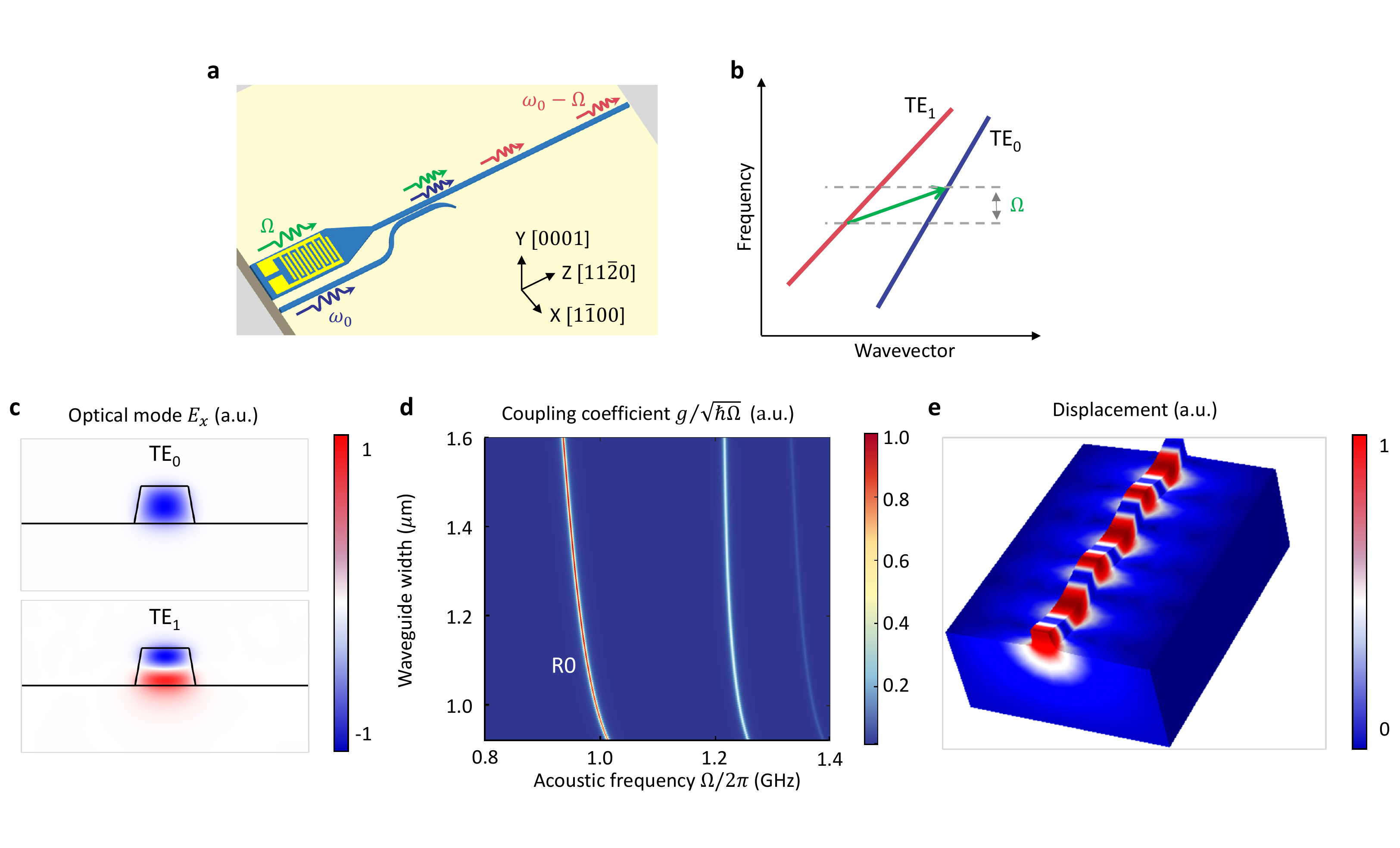}
\caption{GaN-on-sapphire platform for acousto-optic modulation. \textbf{a.} Schematic of the integrated acouto-optic modulator. Waveguides are aligned along the GaN $[11\Bar{2}0]$ direction. \textbf{b.} Phase matching condition for co-propagating acoustic field (green), input TE$_0$ (blue), and output TE$_1$ (red) optical modes. \textbf{c.} Simulated electric field profiles of input TE$_0$ and output TE$_1$ optical modes along the \textit{x} direction. \textbf{d.} Calculated acousto-optic coupling coefficient between TE$_0$ and TE$_1$ optical modes as a function of waveguide width and acoustic frequency. \textbf{e.} Simulated displacement profile of the fundamental Rayleigh (R0) acoustic mode.}
\label{fig:figure1}
\end{figure*}


\section{Results}

The integrated acousto-optic modulator is schematically depicted in Fig.\;\ref{fig:figure1}a. Acoustic fields are launched by interdigital transducers (IDT) through the piezoelectric effect, and focused into the waveguide. Optical fields are launched through a separate waveguide, and transferred into the same waveguide with acoustic fields through a directional coupler.
For co-propagating acoustic and optical fields, AOM can occur as both Stokes and anti-Stokes processes. Input optical fields at angular frequency $\omega_0$ are scattered by acoustic fields at angular frequency $\Omega$ to generate the output optical field at $\omega_1=\omega_0-\Omega$ and $\omega_1=\omega_0+\Omega$ in the Stokes and anti-Stokes processes respectively. 
The corresponding phase matching conditions are $\beta_0 - q=\beta_1$ and $\beta_0 + q=\beta_1$ with $q$, $\beta_0$, and $\beta_1$ the acoustic, optical input, and optical output wavevectors respectively (Fig.\;\ref{fig:figure1}b).
We can switch between Stokes and anti-Stokes processes by interchanging the optical input and output modes~\cite{kittlaus2021electrically}.
Here, we use the fundamental and first-order transverse-electric (TE$_0$ and TE$_1$) modes for the input and output optical fields respectively (Fig.\;\ref{fig:figure1}c). As the TE$_0$ mode has a larger wave-vector ($\beta_0>\beta_1$), the Stokes process dominates the acousto-optic modulation in our device. The conversion between TE$_0$ and TE$_1$ optical modes can be mediated by different acoustic modes. 
We perform numerical simulations to calculate the acousto-optic coupling coefficient~\cite{wiederhecker2019brillouin}. Multiple acoustic modes can mediate efficient acousto-optic coupling between TE$_0$ and TE$_1$ modes as shown in Fig.\;\ref{fig:figure1}d.  We choose the fundamental Rayleigh (R0) acoustic mode, which shows the largest coupling strength. Moreover, with the significant out-of-plane displacement, the fundamental Rayleigh mode can be efficiently excited by IDTs on GaN (0001) plane~(Fig.\;\ref{fig:figure1}e)~\cite{xu2022high}. 
Due to the strong sub-wavelength confinement, the acousto-optic modulation process shows significant geometric dispersion. Therefore, the acoustic frequency can be tuned by the waveguide width. The waveguide width of our device is designed to be 1\;$\mu$m. Therefore, the phase matching condition can be satisfied near the acoustic frequency $\Omega=2\pi\,\times\,$1\;GHz. 
\begin{figure}[htbp]
\centering
\includegraphics[width=7cm]{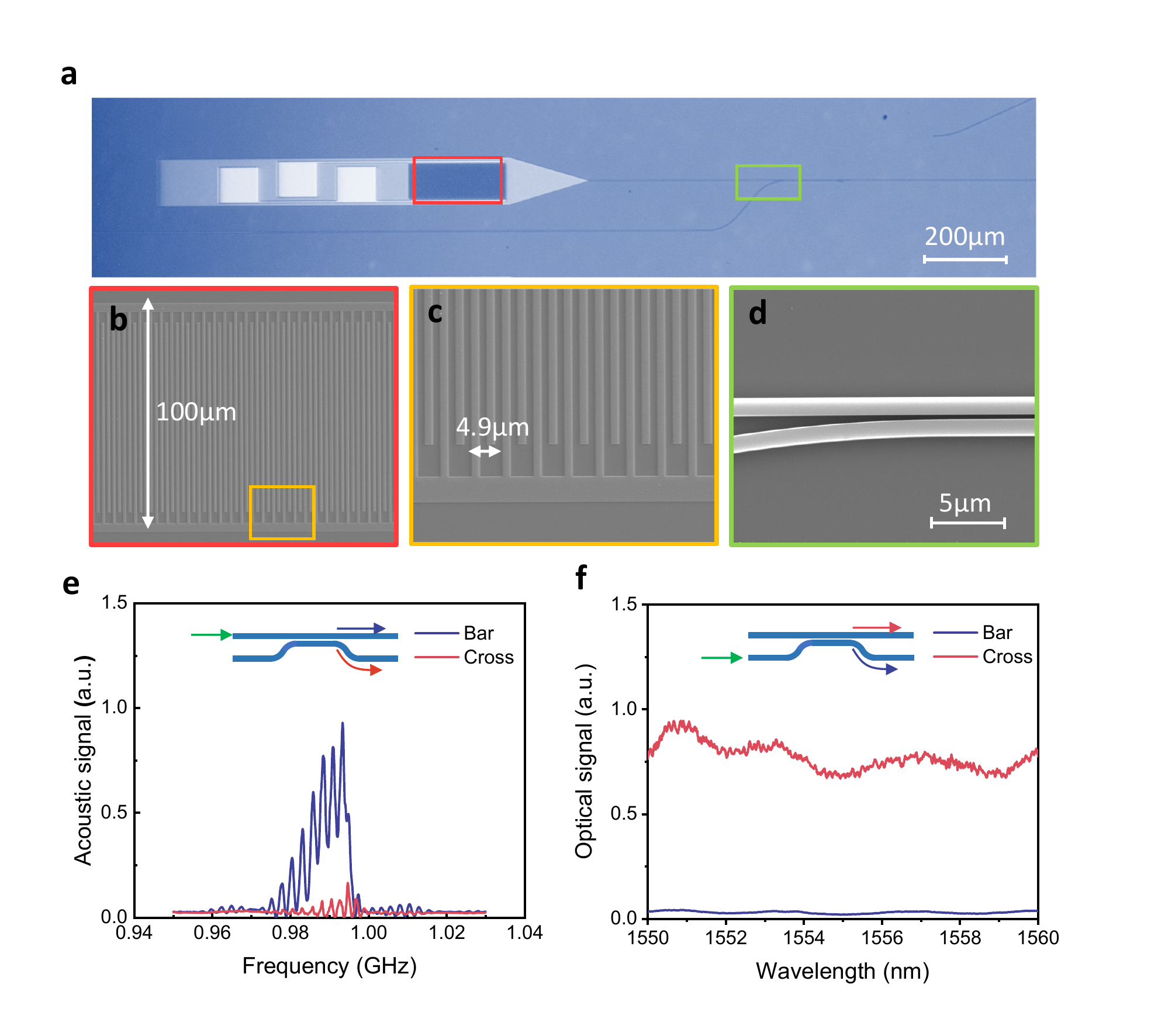}
\caption{Integrated acousto-optic modulators. \textbf{a.} Optical image of the fabricated device. \textbf{b-d.} Scanning electron microscopy (SEM) images of the IDT (\textbf{b}), IDT electrodes (\textbf{c}), and directional coupler (\textbf{d}). \textbf{e.} Acoustic transmission of the directional coupler in the bar (blue) and cross (red) waveguides. \textbf{f.} Optical transmission of the directional coupler in the bar (blue) and cross (red) waveguides.}
\label{fig:figure2}
\end{figure}

The device is fabricated with 1-$\mu$m thick GaN template wafers grown on sapphire substrates using metal-organic chemical vapor deposition~(Fig.\;\ref{fig:figure2}a). Acousto-optic devices are patterned by the electron-beam lithography (EBL) using FOX-16 resist. After developing in TMAH, we etch the GaN layer with reactive ion etching using Cl$_2$/BCl$_3$/Ar gases. IDTs are defined with EBL in ploymethyl methacrylate (PMMA) resist, followed by Ti/Al/Au deposition and lift-off in acetone.
The total waveguide length is $L=3$\;mm to ensure efficient acousto-optic interaction. IDTs consist of a 5-nm titanium bottom layer, 100-nm aluminium middle layer, and 10-nm gold top layer. The IDT period is designed as 4.9\;$\mu$m to match the acoustic frequency around $\Omega=2\pi\,\times\,$1\;GHz (Supplementary Section 1). The IDT aperture and electrode width are 100\;$\mu$m and 1.22\;$\mu$m respectively (Fig.\;\ref{fig:figure2}b and c). 
The directional coupler consists of two parallel 800\;nm wide waveguides with 400\;nm gap (Fig.\;\ref{fig:figure2}d). As acoustic and optical fields have different coupling strengths between the two waveguides, the same directional coupler structure can be designed to show different functions for acoustic and optical fields. Here, we set the directional coupler length at 250\;$\mu$m, in which case acoustic fields remain in the same waveguide and optical fields are completely transferred into the other waveguide (Supplementary Section 2). 

We first use the vector network analyzer (VNA) to measure the acoustic transmission between IDT pairs directly connected by straight GaN waveguides with different lengths. The dependence of the acoustic transmission on the device length allows us to estimate the propagation loss of the fundamental Rayleigh mode, which is $\alpha_b=0.85$\;dB/mm 
(Supplementary Section 1). By extrapolating the acoustic transmission to zero device length, we can obtain the IDT power efficiency $\eta^2=-44$\;dB. With 50 periods, IDTs show a center frequency of 0.99 GHz and 3-dB bandwidth of 12 MHz. To verify the directional coupler performance, we fabricated test devices with four input/output ports connected all with IDTs for acoustic field characterization or all with optical couplers for optical field characterization (Fig.\;\ref{fig:figure2}e and f). We only observe strong acoustic signal in the bar configuration of the directional coupler (Fig.\;\ref{fig:figure2}e). The oscillation in the acoustic transmission spectrum is caused by the reflection between input and output IDTs. Extinction ratio above 10\;dB between the bar and cross waveguides can be achieved. For optical fields, we only observe strong output in the cross configuration with average extinction ratio above 14\;dB (Fig.\;\ref{fig:figure2}f). This shows that optical fields and acoustic fields can be efficiently combined by the directional coupler.

\begin{figure*}[htbp]%
\centering
\includegraphics[width=\textwidth]{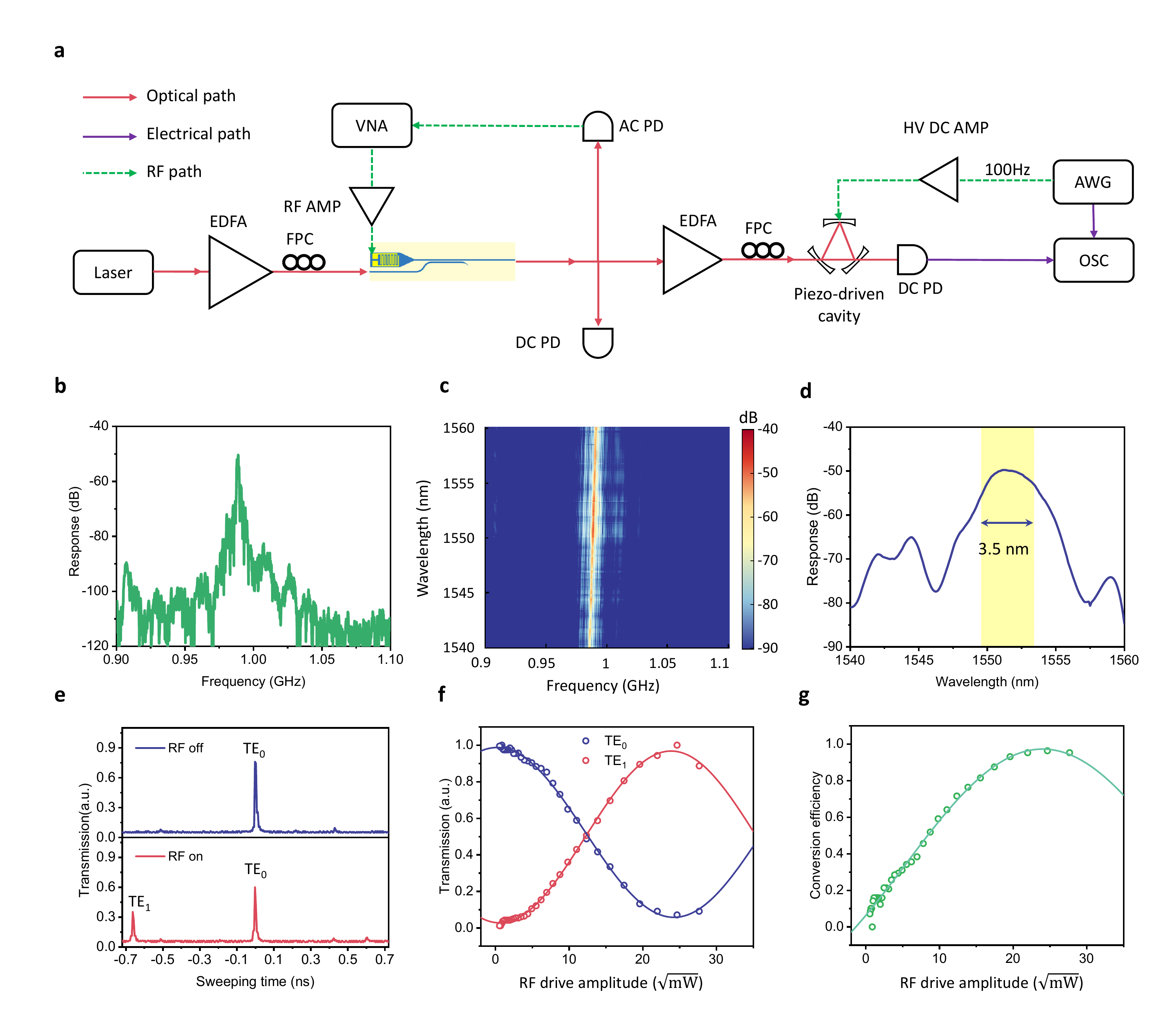}
\caption{Acousto-optic modulation performance. \textbf{a.} Measurement setup. EDFA: Erbium-Doped Fiber Amplifier. FPC: Fiber Polarization Controller. VNA: Vector Network Analyszer. RF AMP: RF Amplifier. HV DC AMP: High voltage DC Amplifier. AWG: Arbitrary Waveform Generator. OSC: Oscilloscope. PD: Photo-Detector.
\textbf{b.} Acoustic modulation spectrum with optical input wavelength 1551.7\;nm. 
\textbf{c.} Acoustic modulation spectrum with different optical input wavelengths. 
\textbf{d.} Optical modulation spectrum with acoustic frequency $\Omega/2\pi =  0.998$\;GHz. 
\textbf{e.} Optical signals filtered by the free-space cavity.
\textbf{f.} Power of TE$_0$ and TE$_1$ modes with different RF driving amplitudes. 
\textbf{g.} Optical conversion efficiency with different RF driving amplitudes.
}
\label{fig:figure3}
\end{figure*}

The modulation response is characterized by the measurement setup illustrated in Fig.\;\ref{fig:figure3}a. CW light generated by a tunable semiconductor laser is
amplified by the erbium-doped fiber amplifier (EDFA) and delivered to the device with a lensed fiber. RF signals from the vector network analyzer (VNA) is loaded into IDTs after amplification to excite acoustic fields. The optical transmission is monitored by a slow high-sensitivity photodetector (DC PD). The output optical field is also measured by a fast photodetector (AC PD), whose electric signal is sent back to VNA. The representative RF modulation spectrum is shown in Fig.\;\ref{fig:figure3}b. The input optical wavelength is set at  $\lambda=$1551.7\;nm. Strong modulation can be clearly observed. The acoustic modulation 3-dB bandwidth is measured as $\Delta$ = 0.57\;MHz. With the modulation bandwidth and acoustic propagation loss, we can calculate the acoustic group velocity $v_{\rm g}=2\pi\Delta/\alpha_b\approx\,$4400\;m/s, which agrees with the simulated value (see Supplementary Section 1)~\cite{liu2019electromechanical}.
We further measure the RF modulation response at different optical wavelengths (Fig.\;\ref{fig:figure3}c). The acoustic frequency with maximum modulation efficiency shifts with the optical wavelength, showing the change of the phase-matching condition due to dispersion. Figure\;\ref{fig:figure3}d shows the peak modulation response at different optical wavelengths with the fixed acoustic frequency $\Omega/2\pi$= 0.998\;GHz. The optical bandwidth of the modulation process is measured as 3.5\;nm, which agrees the theoretical value $\delta\lambda=2.78\lambda^2/(\pi L \delta n_g)\approx3.2$\;nm with $\delta n_{\rm g}=0.22$ the group refractive index difference between TE$_0$ and TE$_1$ modes~\cite{kittlaus2021electrically}.

To measure the optical conversion efficiency, we use a free-space cavity with 1-MHz linewidth as the optical filter (Fig.\;\ref{fig:figure3}a). A piezo-transducer is attached to the free-space cavity mirror to modify the resonant wavelengths. The driving voltage of the piezo-transducer is continuously swept. Therefore, we can separate optical powers from the residual input and converted output optical fields in the time domain (Fig.\;\ref{fig:figure3}e). Without acoustic fields, we only observe the TE$_0$ input optical field. With acoustic fields, the amplitude of the TE$_0$ input optical field decreases. More importantly, we can observe the emergence of the converted TE$_1$ output optical field. The converted TE$_1$ output optical field only shows up on one side of the residual TE$_0$ input optical field, proving the single-sideband nature of acousto-optic modulation. 
By increasing the RF drive amplitude, we can observe the sinusoidal oscillation of the optical power between TE$_0$ and TE$_1$ modes (Fig.\;\ref{fig:figure3}f). With driving power $P=560$\;mW, the TE$_0$ power is close to zero with the TE$_1$ mode reaching the maximum amplitude. Therefore, the on-chip conversion efficiency ($\theta$) close to unity can be achieved at driving power $P=560$\;mW (Fig.\;\ref{fig:figure3}g). This further allows us to estimate the acousto-optic coupling coefficient {(Supplementary Section 3) {~\cite{sarabalis2021acousto}
\begin{equation}
\begin{aligned}
    \frac{g}{\sqrt{\hbar \Omega}} = \frac{\pi\alpha_b}{4\eta\sqrt{P}(1-e^{-\alpha_b L/2})}\approx255\;\mathrm{mm^{-1}W^{-1/2}}
\end{aligned}
\label{eq:sp}
\end{equation}
which is more than two orders of magnitude higher than acousto-optic devices without simultaneous sub-wavelength acoustic and optical confinement~(Table\;1).

\begin{figure}[htbp]
\centering
\includegraphics[width=7cm]{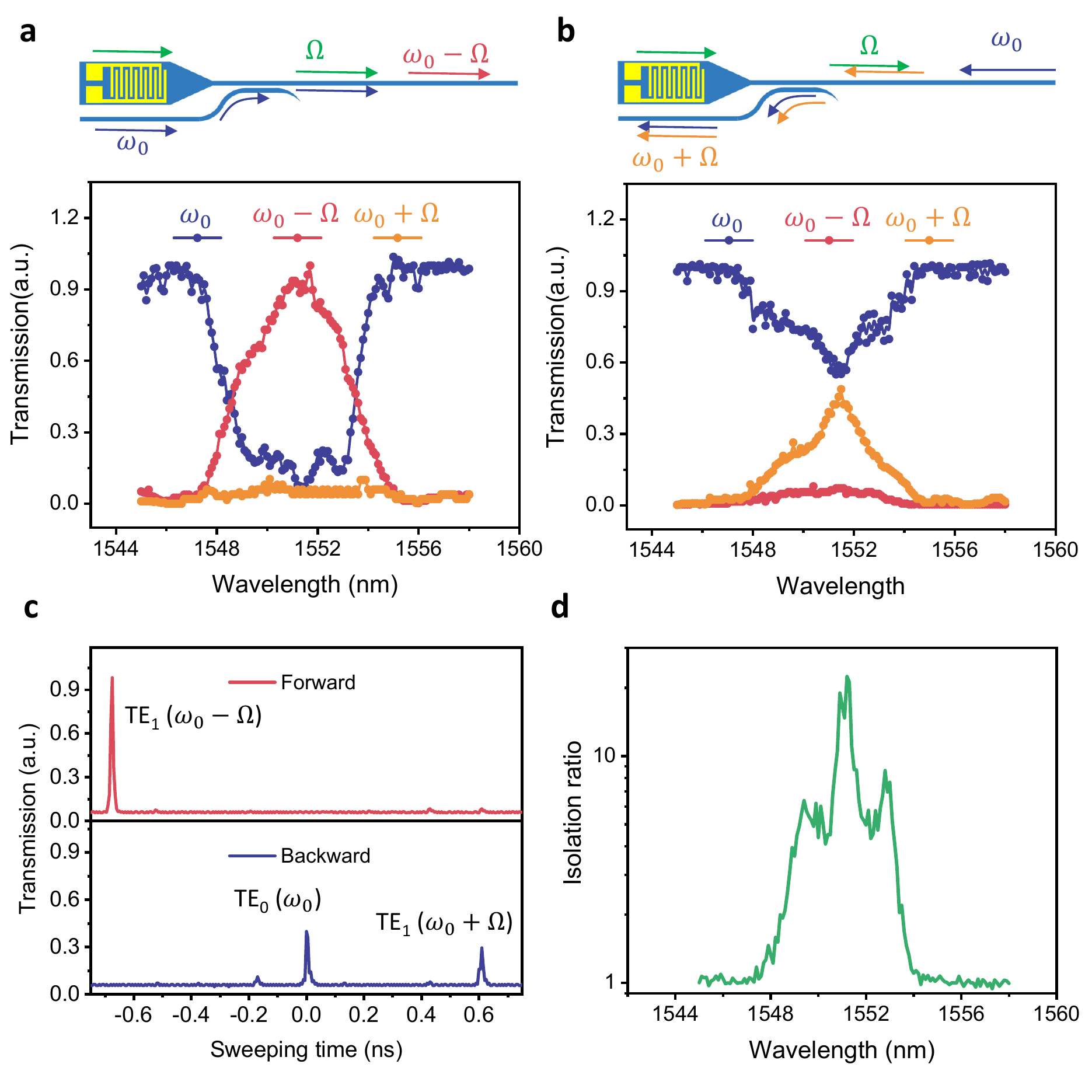}
\caption{Nonreciprocal propagation. 
\textbf{a.} Power spectrum of the input TE$_0$ mode and output Stokes and anti-Stokes TE$_1$ modes in the forward direction.
\textbf{b.} Power spectrum of the input TE$_0$ mode and output Stokes and anti-Stokes TE$_1$ modes in the backward direction.
\textbf{c.} Optical signals filtered
by the free-space cavity in forward and backward directions. \textbf{d.} Power ratio between forward and back directions.}
\label{fig:figure4}
\end{figure}

Due to difficulties in integrating high-quality magneto-optical materials, it is challenging to build on-chip optical isolators and circulators. With the direction-dependent phase-matching condition, acousto-optic modulation provides the non-magnetic method to realize on-chip optical isolation~\cite{kittlaus2018non,kittlaus2021electrically,sarabalis2021acousto}. We verify the non-reciprocal behavior of our acousto-optic device by measuring the optical transmission in both forward and backward directions (Fig.\;\ref{fig:figure4}a and b). The RF driving signal has fixed frequency $\Omega/2\pi=0.998$\;GHz and power $P=560$\;mW.
In the forward direction, phase-matching condition is satisfied for the Stokes process, where the input TE$_0$ mode at frequency $\omega_0$ is converted into the output TE$_1$ mode at frequency $\omega_0-\Omega$  (Fig.\;\ref{fig:figure4}a)~\cite{sarabalis2021acousto}. No anti-Stokes signal at frequency $\omega_0+\Omega$ is observed. 
In the backward direction, no mode conversion should happen ideally and all optical power should remain in the TE$_0$ mode at frequency $\omega_0$. However, with the finite interaction length, we observe the anti-Stokes process in the backward direction (Fig.\;\ref{fig:figure4}b). The anti-Stokes process in the backward direction has much lower efficiency due to phase mismatch. Therefore, the input TE$_0$ mode at frequency $\omega_0$ is only partially converted into the output TE$_1$ mode at frequency $\omega_0+\Omega$ even at the maximum modulation optical wavelength $\lambda=1551.7$\;nm (Fig.\;\ref{fig:figure4}c). If we compare the TE$_0$ power in forward and backward directions, we can clearly see that isolation ratio above 10\;dB has been achieved (Fig.\;\ref{fig:figure4}d).


\section{Discussion}
To benchmark the performance of our device, a summary of recent works on integrated acousto-optic modulators is presented in Table\;\ref{summary}. The GaN-on-sapphire platform in this work is the only one that is capable of confining both optical and acoustic fields in sub-wavelength scales without using suspended structures. Therefore, the interaction length has been significantly improved while maintaining the high acousto-optic coupling coefficient. In addition, the GaN-on-sapphire platform also has excellent power-handling capability. No performance degradation is observed under high driving powers (Supplementary Section 4). Moreover, the GaN-on-sapphire platform shows remarkably lower acoustic propagation loss. The excellent performance in all critical merits leads to the demonstration of the near-unity optical conversion efficiency. The robustness of unsuspended structures also allows the incorporation of integrated acousto-optic modulators into large-scale photonic-phononic circuits with complex functionalities. The elimination of supporting tethers and membranes to anchor suspended structures further enables the flexible design of circuit patterns.

The acoustic driving efficiency can be further improved. Propagation loss below 0.05\;dB/mm has been demonstrated with GaN-on-sapphire acoustic waveguides~\cite{fu2019phononic}. Therefore, the acousto-optic interaction length can be extended by more than 10 times. This will lead to the reduction of the driving power by more than two orders of magnitude, below 10\;mW with the current IDT design.
The IDT efficiency can be further improved using uni-directional IDT designs with more periods and better electrical impedance matching~\cite{yamanouchi2005ultra,rathod2019review}. The heterogeneous structure of aluminum nitride (AlN) and GaN, which is widely used for power electronics~\cite{hamza2020review}, can further increase the driving efficiency by leveraging the larger piezoelectric coefficient in AlN~\cite{rais2014gallium}. As a result, we expect the driving power to achieve complete optical conversion can be decreased to the micro-watt level.

\begin{table*}[htbp]
\centering
\caption{Integrated acousto-optic modulators}
\label{summary}
\resizebox{\textwidth}{!}{
\begin{tabular}{@{}ccccccccccc@{}}
\hline
Work        & Year & Platform & $\Omega/2\pi$ & $\alpha_b$ & $L$ & $g/\sqrt{\hbar\Omega}$ & $\theta$ & Suspended & Optical  & Acoustic\\
        &  &   & (GHz) & (dB/mm) & (mm)&  (mm$^{-1}$W$^{-1/2}$) & ($\%$) & &confined &confined\\ \hline
Liu~\cite{liu2019electromechanical} & 2019 & AlN & 16.4 & --- & 0.5 & 0.041 & 2.5e-4 & Y & Y & Y \\ \hline
Kittlaus~\cite{kittlaus2021electrically} & 2020 & Si/AlN & 3.11 & --- &0.96& 4.5 & 13.5 & N & Y & N \\ \hline
Christopher~\cite{sarabalis2020acousto} & 2020 & LiNbO$_3$ & 0.67 & --- &0.6& 1.66 & 0.9 & N & Y & N \\ \hline
Shao~\cite{shao2020integrated} & 2020 & LiNbO$_3$ & 2.89 & --- &0.1& 0.004 & 3.5 & N & N & N \\ \hline
Ahmed~\cite{hassanien2021efficient} & 2021 & LiNbO$_3$ & 1.16 & 4 & 0.45& 0.417 & 1 & Y & Y & N \\ \hline
Christopher~\cite{sarabalis2021acousto} & 2021 & LiNbO$_3$ &  0.44 & 11.7 &0.25 & 377 & 18 & Y & Y & Y \\ \hline
Wan~\cite{wan2022highly} & 2022 & LiNbO$_3$ & 0.84 & --- &0.12& 0.035 & 3.2e-3 & N & Y & N \\ \hline
This work & 2023 & GaN & 0.99 & 0.85 & 3& 255 & $\sim$100 & N & Y & Y \\ \hline
\end{tabular}
}
\end{table*}

\section{Conclusion}

In conclusion, we have developed the GaN-on-sapphire platform for acousto-optic devices. We achieve high acousto-optic coupling strength, long interaction length, low-loss acoustic propagation, and large power handling capability at the same time. This leads to the first demonstration of near-unity conversion efficiency with integrated acousto-optic modulators. This work will enable the exploration of hybrid photonic-phoninc circuits at large scale for advanced signal processing, with important applications in microwave photonics and quantum transduction.





\begin{backmatter}

\bmsection{Funding}
This material is based upon work supported by the Office of the Under Secretary of Defense for Research and Engineering under DEPSCoR program award number FA9550-21-1-0225 managed by Army Research Office, and NSF Grant No. ITE-2134830.


\bmsection{Disclosures}
The Authors declare no competing financial or non-financial interests



\bmsection{Data availability} The data that support the findings of this study are available from the corresponding author upon reasonable request

\bigskip



\bigskip


\bmsection{Supplemental document}
See Supplement 1 for supporting content. 

\end{backmatter}


\bibliography{Ref}






\end{document}